\documentclass[12pt]{iopart}
\usepackage{graphicx}
\usepackage{iopams}

\def\Journal#1#2#3#4{{#1}{\bf #2}, #3 (#4)}

\def\PRL{Phys. Rev. Lett.\ }

\def\PLB{{Phys. Lett.}~{\bf B}}

\def\NIMA{{Nucl. Instrum. Methods}~{\bf A}}
\def\ZPA{{Z. Phys.}~{\bf A}}

\begin{document}

\title{Search for the 
$\overline{\Theta}^- \rightarrow $ K$^-$\,$\overline{n}$ with PHENIX}

\author{Christopher Pinkenburg$^{\dag}$~ for the PHENIX\footnote[3]{For the 
full PHENIX Collaboration author list and acknowledgments , see 
Appendix ``Collaborations'' of this volume.} collaboration}
\address{\dag Brookhaven National Laboratory, Physics Dept., Upton, NY, 11973-5000}

\ead{pinkenburg@bnl.gov}

\begin{abstract}
The PHENIX experiment at RHIC should be 
sensitive to decays of the the anti--pentaquark $\overline{\Theta}^-$
via the K$^-$\,$\overline{n}$ channel. 
Charged kaons can be identified using the standard tracking
and time of flight up to a momentum of 1.5 GeV/c. Anti--neutron 
candidates are detected via their 
annihilation signal in the highly segmented electromagnetic 
calorimeter (EMCal). In order 
to assess the quality of the anti--neutron identification we reconstruct the 
$\overline{\Sigma} \rightarrow \overline{n}\pi$. As an additional crosscheck
the invariant mass of K$^+$\,$\overline{n}$ is reconstructed where no resonance 
in the pentaquark mass range is expected. At the present time no enhancement at the
expected pentaquark mass is observed in dAu collisions at $\sqrt{s_{NN}}~=~$200\,GeV.

\end{abstract}

The possibility of five quark systems (pentaquarks)  has been discussed for 
more than two decades (see, e.g. Ref~\cite{Jaffe:1976ii}). 
A recent publication \cite{Diakonov:1997mm}
based on the soliton model made a prediction of a narrow resonance 
($\Gamma$$<$15MeV/c$^2$)
with a mass $\approx$ 1530 MeV/c$^2$. This pentaquark -- designated the 
$\Theta^+$ -- is expected to decay
into the channels K$^+$\,n and K$^0$\,p. Starting in 2003 narrow resonances in
the expected mass range have been observed in many
experiments (LEPS \cite{Nakano:2003qx}, DIANA \cite{Barmin:2003vv},
CLAS \cite{Kubarovsky:2003fi} \cite{Stepanyan:2003qr}, SAPHIR \cite{Barth:ja}
HERMES \cite{Airapetian:2003ri},
SVD \cite{Aleev:2004sa}
COSY-TOF \cite{Abdel-Bary:2004ts}). A peak was also found in neutrino 
scattering experiments \cite{Asratyan:2003cb}. 

The reconstruction of the $\Theta^+$ pentaquark is technically difficult in 
PHENIX \cite{nim_phenix}, due to the relatively small acceptance for 3-body 
final states and the difficulty of detecting neutrons. 
However, due to the unique signature of anti--neutrons in the highly 
segmented PHENIX electromagnetic calorimeter, a search for decays of 
the anti--pentaquark  
$\overline{\Theta}^- \rightarrow $ K$^-$\,$\overline{n}$ is technically feasible. 

\begin{figure}
\center
\includegraphics[width=6 cm]{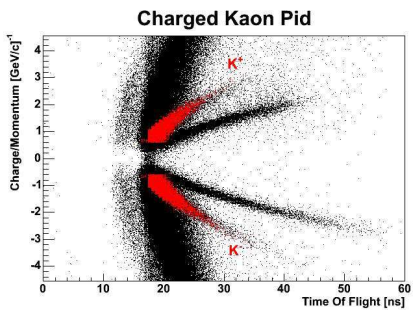}
\caption{Charged kaon identification: The momentum is reconstructed by the central tracking, the
time-of-flight is determined by the EMCal. The range of identified kaons is marked in black.}
\label{fig:momtof}
\end{figure}

Charged particles are tracked using the central arm spectrometers 
\cite{PHENIX_tracking}. The kaon identification is
accomplished by combining their momentum obtained from the tracking detectors 
with their time-of-flight as measured by the EMCal, shown in 
fig.\,\ref{fig:momtof}. The upper limit of the momenta for separating kaons 
is 1.5 GeV/c, beyond which the contamination by pions becomes too large. 

\begin{figure}
\center
\includegraphics[width=9 cm]{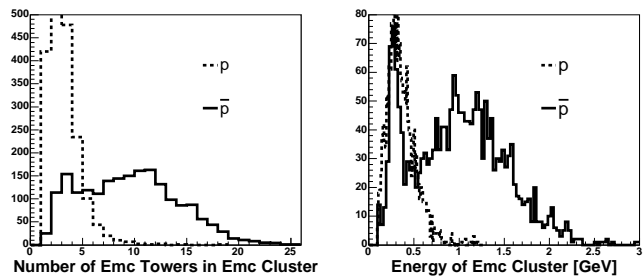}
\caption{EMCal response for protons and anti--protons. The additional annihilation energy of the
anti--protons leads to larger clusters. Searching for large clusters is the main tool to identify
anti--neutron candidates}
\label{fig:emctwrs}
\end{figure}

The anti--neutron candidates are selected via their annihilation
signal in the EMCal. Since there is no independent measurement of 
anti--neutrons to calibrate the EMCal response, 
guidance for identifying the anti--neutron signal is provided by the 
characteristics of clusters created by identified protons and anti--protons.
The identification of both protons and anti--protons is accomplished via the 
usual combination of momentum and time-of-flight, and the resulting sample is 
used to determine the features of  the annihilation signal in the calorimeter. 
The main differences in the response of the EMCal to protons and anti--protons are the number of 
struck towers and the amount of energy deposited, as shown in fig.\,\ref{fig:emctwrs}.  Only clusters with a measured 
time more than 3\,ns later than the photon arrival time were used in the analysis, and further, clusters were 
only used if the shower shape showed a poor fit to that expected for a photon.

To remove clusters which were produced by charged particles a layer of pad 
chambers (PC3) in front of the EMCal was used as a veto counter. 
Clusters within 12\,cm of a PC3 hit were excluded. In order to compensate for 
dead regions of the PC3, EMCal clusters which are closer than 12\,cm to 
the trajectory of a charged track at the EMCal surface were also removed. 

\begin{figure}
\center
\includegraphics[width=8.5 cm]{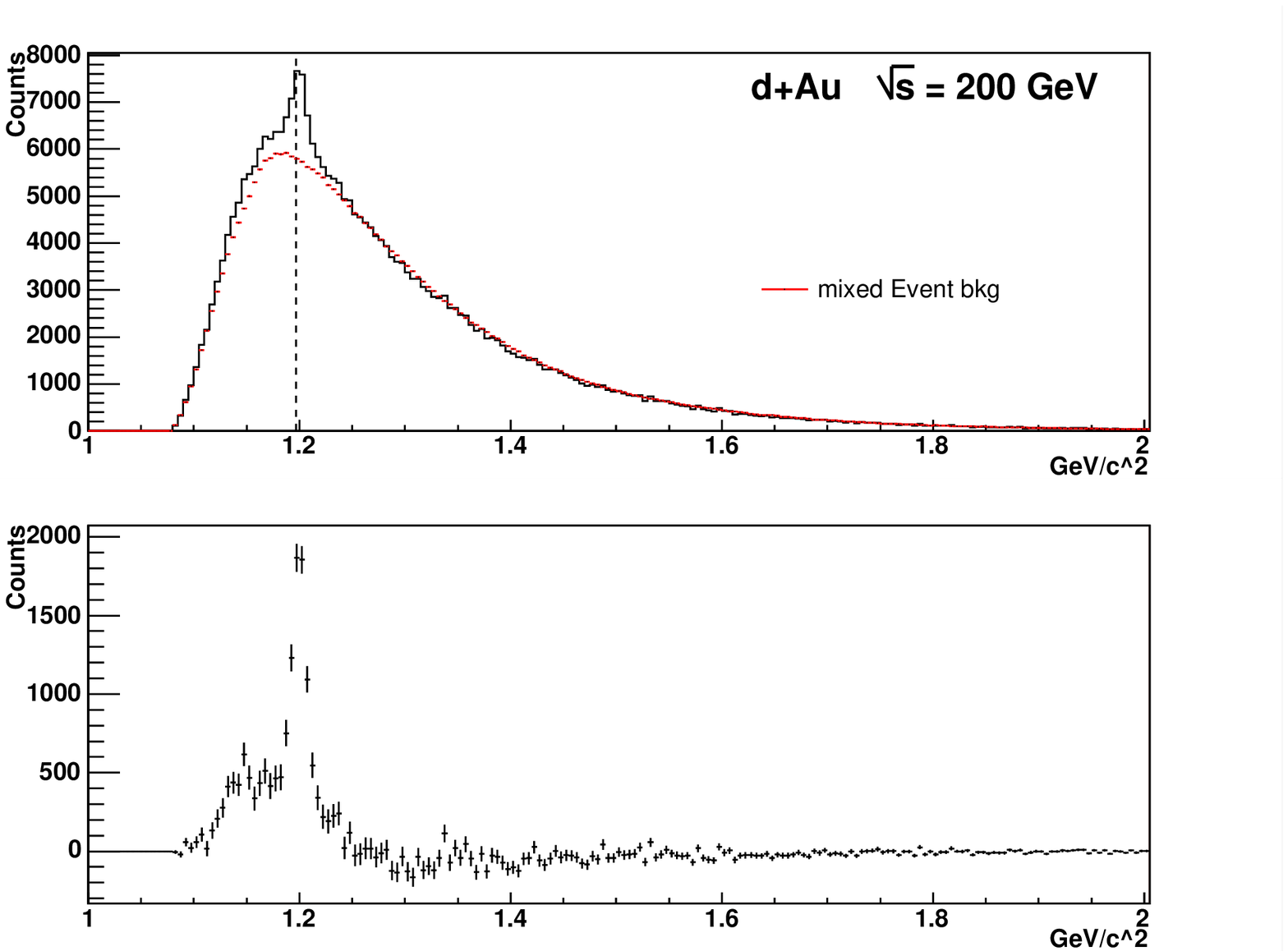}
\caption{$\pi^+$\,$\overline{n}$ Invariant mass distribution. The peak is very close to the expected value of 
1.197 GeV/c$^2$. The lower panel shows the distribution after the mixed background is subtracted. This plot contains the timing correction. Without the
timing correction the width 
of the peak is about a factor of 2 larger.}

\label{fig:sigplus}
\end{figure}

The selected anti--neutron candidates are then combined with identified charged kaons (and pions for the reconstruction
of $\overline{\Sigma}$ baryons) and the invariant mass of these pairs is calculated. To subtract the uncorrelated background a 
high statistics distribution using pairs from mixed events is created.
The reconstruction of anti--sigmas provides an important confirmation of the validity of these techniques. 
As shown in fig.\,\ref{fig:sigplus},
the $\pi^+$\,$\overline{n}$ invariant mass distribution 
exhibits a peak very close to the nominal mass of the $\overline{\Sigma}^-$ (1.197 GeV/c$^2$ \cite{pdg}).
Future PHENIX limits or yields for  anti--pentaquark production at RHIC are likely to rely on this observation
as an important calibration of both experimental resolution and acceptance for anti--neutrons. 

\begin{figure}
\center
\includegraphics[width=8.5 cm]{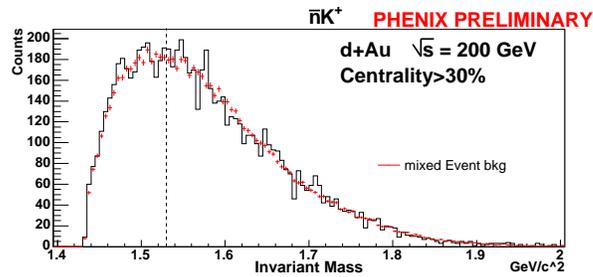}
\caption{K$^+$\,$\overline{n}$ Invariant mass distribution. No peak is expected in
the mass range between 1.5\,GeV/c$^2$ and 1.6\,GeV/c$^2$. It serves as
a crosscheck for technical problems and particle misidentifications.}
\label{fig:kplusnbar}
\end{figure}

A further crosscheck for a real anti--pentaquark signal is the absence of a signal in the invariant mass distribution 
of K$^+$\,$\overline{n}$ which is shown in fig.\,\ref{fig:kplusnbar}. 
No resonance is expected
in the mass range of 1.5GeV/c$^2$ -- 1.6GeV/c$^2$ but technical problems and contaminations due to misidentified particles
should affect both distributions.

The analysis was done with d\,+\,Au data at $\sqrt{s_{NN}}~=~$200\,GeV which were taken during Run-3 at RHIC. 
In the initial analysis of 
peripheral events (Centrality$>$30\% of the inelastic cross section) a statistically 
significant peak at an invariant
 mass of 1.54\,GeV/c$^2$ was observed but no peak was seen in the accompanying 
K$^+$\,$\overline{n}$ invariant
mass distribution. This was the status of the analysis which was reported at the conference.
As part of a systematic investigation of this intriguing result, an independent analysis 
was performed which showed no structure in the vicinity of 1.54\,GeV/c$^2$. 
Further comparisons determined that the original analysis
lacked a necessary timing correction. This translated into a distortion 
of the momentum of only the 
anti--neutrons.

\begin{figure}
\center
\includegraphics[width=8.5 cm]{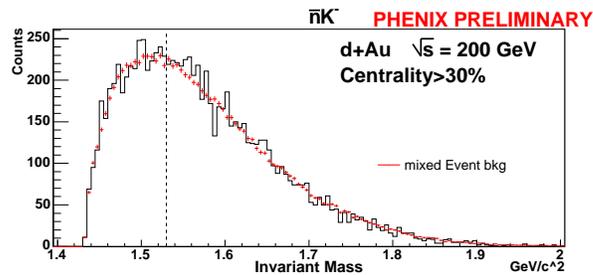}
\caption{K$^-$\,$\overline{n}$ Invariant mass distribution, no enhancement is visible 
at the $\overline{\Theta}^-$ mass after the necessary timing correction is applied}
\label{fig:theta}
\end{figure}

The correction is negligible ($<$5\%) for most of the events, 
which is the reason why the original analysis could reconstruct the $\overline{\Sigma}$. 
However, as a result of applying the
correction the peak at 1.54 GeV/c$^2$ in the K$^-$\,$\overline{n}$ invariant mass distribution
loses its statistical significance. The distribution shown in  fig.\,\ref{fig:theta}
uses the same events and the same cuts as those
originally presented, but after application of the correction. The K$^+$\,$\overline{n}$ invariant mass distribution in fig.\,\ref{fig:kplusnbar} (also after
application of the correction) which served as a safeguard against 
technical problems and particle misidentifications 
did not change visibly.

Currently it is unclear what mechanism is behind the 
appearance of the peak at 1.54\,GeV/c$^2$ and why the control K$^+$\,$\overline{n}$ invariant mass 
distribution did not exhibit the same feature. This is being actively
investigated. The unique anti--neutron capabilities of the PHENIX apparatus, as evidenced by the
cleanly reconstructed anti--sigma channel, leave open the possibility 
of  testing of coalescence models for their
production\cite{Chen:2003tn,Randrup:2003fq}
by establishing limits
or measuring yields of anti--pentaquarks at RHIC.

\end{document}